\documentclass[final,5p,times,twocolumn,dvipsnames]{elsarticle}

\usepackage{amsmath,amssymb}  % For mathematical symbols
\usepackage{graphicx}         % For including graphics
\usepackage{hyperref}         % For hyperlinks
\usepackage{xcolor,here}
\usepackage{lineno}           % For line numbers (optional)
\usepackage{multicol}         % For multi-column formatting
\usepackage{tikz}
\usetikzlibrary{shapes,arrows,calc}

\newcommand{\e}{\vskip 0.5mm \ \\}

\journal{Automatica}

\begin{document}

\begin{frontmatter}

% Title
\title{On reconstructing high derivatives of noisy time-series with confidence intervals}

% Authors and Affiliations
\author[gipsa]{Mazen Alamir\corref{cor1}}

\cortext[cor1]{Corresponding author}
\ead{mazen.alamir@grenoble-inp.fr}
% Affiliations
\address[gipsa]{Univ. Grenoble Alpes, CNRS, Grenoble INP, GIPSA-lab, 38000 Grenoble, France}

% Abstract
\begin{abstract}
    {  Reconstructing high derivatives of noisy measurements is an important step in many control, identification and diagnosis problems. 
    In this paper, a heuristic is proposed to address this challenging issue. The framework is based on a dictionary of identified models indexed by the bandwidth, the noise level and the required degrees of derivation. Each model in the dictionary is identified via cross-validation using tailored learning data. It is also shown that the proposed approach provides heuristically defined confidence intervals on the resulting estimation.  The performance of the framework is compared to the state-of-the-art available algorithms showing noticeably higher accuracy. Although the results are shown for up to the $4$-th derivative, higher derivation orders can be used with comparable results.}
\end{abstract}

% Keywords
\begin{keyword}
    High Derivatives Reconstruction \sep Noisy Time-Series \sep Confidence Interval \sep  Cross-Validation \sep Algorithms.
\end{keyword}

\end{frontmatter}

% Main Text
\section{Introduction}
Reconstructing  high derivatives of measurements is a key issue in some relevant dynamic systems-related topics such as system identification \cite{ljung2010perspectives}, state/parameter estimation \cite{evensen2022data} and anomaly detection to cite but few examples. When dealing with this important issue, two kinds of paradigms should be formally distinguished, namely: 1) the filtering paradigm where a real-time updating of the estimated derivatives is operated, generally via simple difference equations implementing short memory filters and 2) the derivatives reconstruction paradigm that can be done off-line and might involve a more holistic view of the past measurements {  At the price of extra-computational cost.}
\e Indeed, in the first case, new measurements are coming in a real-time stream and a processing algorithm has to take them into account \textit{on the fly} in some updating iteration such as in Kalman Filtering (KF) \cite{khodarahmi2023review} so that the result can feed a control algorithm for instance. The excellent survey \cite{mojallizadeh2021time} provides a recent survey of the state of the art on real-time differentiation algorithms including sliding modes \cite{levant2018sliding}, Kalman Filter and some other few alternatives.
\e In the second case, one disposes of collected measurements, over some observation window, and can process them in order to deliver reconstructed profiles of $d$-th derivatives. In this context, the reconstructed derivatives might serve in an off-line task such as nonlinear continuous-time identification \cite{brunton2022data} and or characterization of normality space \cite{bergweiler2001normality, de2012use}. Spectral Derivation (SD) \cite{davies1995spectral} is an example of such a solution  since the computation of the Fourier transform needs a large measurement window to be processed. {  Obviously, there is no neat separation between the above mentioned classes of methods since window-based reconstruction process can be eligible for on-line real-time implementation if the characteristic time of the systems is not too fast.}
% \e 
% As a matter of fact, there is no neat separation \textit{boundary} between the two above cited contexts. For instance, the different versions of the algorithms involved in the so-called Moving-Horizon Estimators (MHE) \cite{rawlings2021moving}, exploit a moving measurement window over which a whole bunch of past measurements are available and can be processed despite the fact that MHEs are used in real-time in order to update the current estimation. Separation boundaries materialize when the processing time of past windows of measurements becomes close to the inherent characteristic time of the dynamical system. 
\e 
The starting point of this contribution lies in the recent results reported in \cite{mojallizadeh2021time} regarding the \textit{filtering}-like algorithms. These results suggest that, relying on such filtering algorithms to reconstruct higher derivatives (derivation orders $\ge 2$) is not realistic even for quite low noise ($< 3\%$). On the other hand, the emergence of \texttt{Machine Learning} culture where learning data-sets are collected and then processed off-line in order to fit various models and to gain some measurement-enforced understanding of hidden relationships triggered an interest in the second paradigm mentioned above. The framework proposed in this paper is the consequence of this fact. 
\e
In a nutshell, the framework is based on a dictionary of identified models indexed by the bandwidth, the noise level and the required degrees of derivation. Each model in the dictionary is identified via cross-validation using tailored learning data. In the exploitation mode, the bandwidth and the noise-level are identified from the specific noisy time-series before the derivatives are reconstructed using the specific element in the dictionary of models. The effectiveness of the algorithm comes from the fact that several estimations are obtained for each sampling instant thanks to a moving-window process. This adds to the possibility of getting high reconstruction precision the ability to provide confidence intervals on the resulting estimation. 
\e 
{  It is important to underline that, as explained above, the proposed heuristic is not intended to replace the filtering approaches in all situations. Rather, it can be preferred due to its higher precision in situations where real-time computation is not the key challenge. This might be because the dynamic system is quite slow or when off-line use is targeted for a continuous-time identification or modeling tasks to cite but two examples. }
\e 
This paper is organized as follows. First of all, Section \ref{sec:pbstat} clearly states the problem to be addressed. Section \ref{sec:defnot} introduces some definitions and notation used throughout the paper. The proposed high-order derivatives reconstruction framework is explained in Section \ref{sec:proposed}. Finally Section \ref{secs:assessment} proposes some numerical investigations in order to assess the relevance of the proposed framework and to compare it to some other alternatives.
\section{Problem Statement}\label{sec:pbstat}
\noindent The objective of the paper is to provide an algorithm that 
takes as argument a triplet ($\mathbf s,d,\tau)$ where:
\begin{itemize}
\item[$\checkmark$]$\mathbf s$ is a sequence $\mathbf s:=(s_1\dots,s_n)$ containing $n$ successive uniformly distributed measurement instances;
\item[$\checkmark$] $d\in \mathbb N$ is the order of derivation;
\item[$\checkmark$] $\tau>0$ is the sampling measurement acquisition period;
\end{itemize}  
and delivers as output a sequence $\hat{\mathbf s}^{[d]}\in \mathbb R^n$ representing an estimation of the $d$-derivative of the original time-series $\mathbf s$ over the same window of length $n$. Moreover, the algorithm should deliver an associated standard deviation profile $\hat \sigma^{[d]}\in \mathbb R^n$ that enables to reconstruct a confidence tube (commonly defined using $\pm 3\hat\sigma^{[d]}$ around $\hat{\mathbf s}^{[d]}$). This targeted map can be denoted by:
\begin{equation}
(\hat{\mathbf s}^{[d]},\hat \sigma^{[d]}) \leftarrow \texttt{est\_deriv}(\mathbf s, d, \tau).\label{targetedmap}
\end{equation} 
A weaker version might require the caller of the function to provide the standard deviation of the measurement noise, referred to hereafter as the \texttt{noise\_level} , namely:
\begin{equation}
(\hat{\mathbf s}^{[d]},\hat \sigma^{[d]}) \leftarrow \texttt{est\_deriv}(\mathbf s, d, \tau, \texttt{noise\_level}).\label{targetedmap_weaker}
\end{equation}
%In the sequel, the two versions of the algorithm are discussed. 
% \e Obviously, the automated estimation of the \texttt{noise\_level} is more attractive as it 1) enables an adaptive processing through a window-dedicated estimation of the noise level (leading to higher precision) and 2) frees the user from the task of investigating the noise level. 
% \e Nevertheless, this adaptive algorithm comes with a \textit{slightly} heavier computational time that can be decisive when using the derivation process in the construction of large training data as it is the case in some \texttt{Machine Learning}-like applications. In such cases, it might be preferable to use \eqref{targetedmap} if the increment of the quality of the reconstruction does not justify the added induced computational cost.
The following section gives some definitions and notation that are extensively used in the presentation of the proposed computational framework.
\e Notice that since there are as many sampling acquisition periods as there are real-life use-cases, the following straightforward identity is worth recalling:
\begin{equation}
\texttt{est\_deriv}(\mathbf s, d, \tau) \equiv \dfrac{1}{\tau^d}\times \texttt{est\_deriv}(\mathbf s, d, 1) \label{obviousidentity}
\end{equation} 
which comes from the possibility of scaling the time so that the resulting scaled sampling period becomes equal to 1. This simply means that  one can concentrate on the normalized case where $\tau=1$, build the associated map and only when the concrete final estimation is operated using the designed map, one can introduce the sampling period-related correction given by \eqref{obviousidentity}. This is of a tremendous importance since the learning step can be done \textit{universally} once for all independently of the effective acquisition period used in the application instances.  Consequently, in the sequel, the normalized case $\tau=1$ is considered dropping here and there the word \textit{normalized} for the sake of brevity when no ambiguity is possible.
\section{Definitions and notation}\label{sec:defnot}
\noindent {  In the sequel, the following notation is used: Given a matrix $M$, the notation $M[i_0:i_1, j_0:j_1]$ denotes the sub-matrix containing the lines from $i_0$ to $i_1$ and the columns from $j_0$  to $j_1$. When the initial index is absent, this means that all the initial indexes are considered up to $i_1$ or $j_1$. The set of $n$ uniformly distributed values between two bounds $v_1$ and $v_2$ is denoted by \texttt{linspace($v_1, v_2, n$)}. Similarly,  \texttt{logspace($v_1, v_2, n$)} denotes the logarithmic version, namely the set defined by $\{10^{\xi}, \xi\in \texttt{linspace($v_1, v_2, n$)}\}$. Finally, considering a time series $\mathbf s\in \mathbb R^{n}$, the notation $\mathbf s[i_1:i_2]$ refer to the \textit{slice} of $\mathbf s$ indexed by indices from $i_1$ to $i_2$: 
\begin{equation}
\mathbf s[i_1:i_2]:= \begin{bmatrix}
    s_{i_1}&s_{i_1+1}&\dots&s_{i_2}
\end{bmatrix} \label{slicings}
\end{equation} 
}  
Now assuming that \texttt{n\_per\_period} (=5 in the implementation) samples are considered to be necessary inside a period to reconstruct a sinusoidal signal, the following maximum normalized pulsation is defined:
\begin{equation}
\text{(Max normalized pulsation)}\qquad \bar \omega:= \dfrac{2\pi}{\texttt{n\_per\_period}}\label{defofwbar}
\end{equation} 
This enables to define a dense grid of pulsations $\bar \Omega_\texttt{grid}\in \mathbb R^{n_\texttt{grid}}$ ($n_\texttt{grid}=200$ is used hereafter): 
\begin{equation}
\bar \Omega_\texttt{grid}:= \{\Omega_1,\dots,\Omega_{n_\texttt{grid}}\}:=\texttt{logspace(-3,0,$n_\texttt{grid}$)}\times \bar \omega\label{defdeOmegagrid}
\end{equation} 
that are used to generate the learning data by randomly drawing a high number of samples representing each a different linear combination of sinusoidal signals with pulsations included in $\bar \Omega_\texttt{grid}$. 
% When a high number of such instances of time-series is generated, it is expected that the resulting profiles are representative enough to any possible smooth time-series within the  allowed bandwidth range (bounded by $\bar\omega$) so that a learning process based on the resulting learning dataset provides a model that correctly generalizes over new unseen instances. 
\e More precisely, given a sequence length $n_t$, one can construct the basis function matrices with columns inside the following set ($t\in \mathbb R^{n_t}$ is the time vector)
\begin{equation}
\Bigl\{\mathbf 1\Bigr\}\cup \Bigl\{\sin(\omega t), \cos(\omega t)\Bigr\}_{\omega\in \bar\Omega_\texttt{grid}} \label{lespulsss}
\end{equation}
leading to a basis function matrix \texttt{B}  such that:
\begin{equation}
\qquad \texttt{B}\in \mathbb R^{n_t\times n_B}\quad\text{where}\quad  n_B:=(2n_\texttt{grid}+1)
\end{equation} 
such that for each randomly sampled vector of coefficients $a\in \mathbb R^{n_B}$, a candidate \textit{admissible} time-series $\texttt{B}a\in \mathbb R^{n_t}$ is obtained. 
\e 
Similarly, using the $d$ derivative of the above defined columns, one defines the $d$-derivative basis functions, denoted hereafter by \texttt{B}$_d$ so that for any $a\in \mathbb R^{n_B}$, the following holds true:
\begin{equation}
\bigl[\texttt{B}_d\bigr] a \quad \text{is the $d$-th derivative of $\bigl[\texttt{B}\bigr]a$ } 
\end{equation} 
Notice that the $n_B$ columns of \texttt{B} correspond to increasing pulsations $1,\Omega_1,\Omega_1,\Omega_2,\Omega_2,\dots$ that are inherited from the pulsations included in $\Omega_\texttt{grid}$ given by \eqref{lespulsss}. Therefore giving a prescribed cut-of pulsation $\Omega\le \bar\omega$, the sub-matrix obtained from \texttt{B} by keeping only the columns corresponding to pulsations that are lower than $\Omega$ is denoted hereafter by \texttt{B[:,:$\Omega$]} representing the (Bandwidth limited Basis):
\begin{equation}
\texttt{B[:,:$\Omega$]} \quad \text{Only columns with pulsations $\le \Omega$}\label{defdeBpointpointomega}
\end{equation} 
This sub-matrix can then be used to compute the projection of any time-series $\mathbf s\in \mathbb R^{n_t}$ on the sub-space of time-series of bandwidth lower than $\Omega$ according to:
\begin{equation}
\hat{\mathbf{s}}(\Omega):=\Pi(\Omega)\cdot \mathbf s\quad \text{where}
\quad {\color{Black}\Pi(\Omega):=\texttt{B[:,:$\Omega$]}\texttt{B[:,:$\Omega$]}^\dag}
\end{equation} 
More generally, since the length of the time-series to be processed is not necessary of length $n_t$, the same projection process can be adapted to any time-series $\mathbf s\in \mathbb R^n$ of length $n\le n_t$ by selecting the first $n$ lines of the matrix \texttt{B} which are denoted hereafter by \texttt{B[1:n,:$\Omega$]} leading to the associated projection matrix:
\begin{equation}
\Pi_n(\Omega):= \texttt{B[1:n,:$\Omega$]}\texttt{B[1:n,:$\Omega$]}^\dag\label{defdePin}
\end{equation}
As mentioned earlier, it is a fact that the appropriate parameters of any differentiation algorithm heavily depend on the frequency content (the bandwidth) of the time-series on the considered observation window. This is the reason why a set of $n_r$ models are identified hereafter for different $n_r$ predefined design cut-of pulsations that belong to $\bar\Omega_\texttt{grid}$. For obvious memory reasons, the number $n_r$ of design pulsations is $\ll n_\texttt{grid}$. More precisely, the following set is defined ($n_r=21$ is used in the implementation):
\begin{equation}
\bar \Omega_\texttt{design}:= \{\omega_1,\dots,\omega_{n_r}\}:=\texttt{linspace($\Omega_1$,$\bar\omega$,$n_r$)}\label{defdeOmegadesign}
\end{equation} 
in which $n_r$ values ranging from the minimal to the maximal values in $\bar\Omega_\texttt{grid}$ are used.
\e 
So far, we have all that we need to describe the proposed algorithm. This is done in the following section. 
\section{The proposed derivation framework}\label{sec:proposed}
\noindent The proposed framework is based on a sequence of ideas. In this section,  each of these ideas is stated, then discussed before its algorithmic translation is given. 
\subsection{The main ideas and their implementation}
\noindent In what follows, the notation $\mathbf y\in \mathbb R^{n_w}$ is used to denote time-series of a specific length $n_w$ which is the dimension of the input space of the maps to be identified. These maps are then used to construct the $d$-derivatives of any time series $\mathbf s$ of length greater than $n_w$ as shown later on. The first idea can be stated as follows:
\begin{center}
\begin{tikzpicture}
\node[rounded corners, fill=Black!5, inner sep=4mm](O){
\begin{minipage}{0.44\textwidth}
\small 
\color{Black}
Given a time-series $\mathbf y\in \mathbb R^{n_w}$, the map $F^{(d)}:\mathbb R^{n_w}\rightarrow\mathbb R^{n_w}$ that estimates the $d$-th derivative of $\mathbf y$ is a linear map, namely $$F^{(d)}(\mathbf y)=A\mathbf y.$$ The matrix $A$ of this map depends on:\vskip 1mm
\begin{enumerate}
    \item the bandwidth of $\mathbf y$ and 
    \item the \texttt{noise\_level} (standard deviation) of the measurement noise. 
\end{enumerate}
\end{minipage} 
};
\node[above, anchor=south west] at(O.north west){\sc Idea 1: A Differentiator is a linear map that is not universal};
\end{tikzpicture}
\end{center} 
Discussion. {\textit{This idea comes from the fact that in the absence of noise, one can imagine a linear interpolation process of sufficiently high order over a functional basis of known derivatives which involves only linear operations in the input $\mathbf y$. In the presence of noise, the appropriate regularization that corresponds to a trade-off enabling tracking the signal without tracking the noise obviously depends on both the noise level and the dynamics of the system since both these characteristics affect the optimal tuning of the trade-off}.
\e 
\textbf{Implementation.} Since the model depends on the bandwidth and the level of the measurement noise, let us consider the set $\bar\Omega_\text{design}$ of Design Pulsations defined in \eqref{defdeOmegadesign} and indexed by the set of integers $j\in \{1,\dots,n_r\}$ together with the following set of \textit{normalized}\footnote{In the sense that the corresponding measurement noise is added to normalized time-series to get time-series of amplitudes lower than 1) to build the learning data.} noise levels given by:
\begin{equation}
\mathcal N:=\Bigl\{\nu_1,\dots,\nu_q\Bigr\}\subset [0.0,0.01,\dots,0.25] \label{noiselevels}
\end{equation} 
indexed by the integer $\ell\in \{1,\dots,q=21\}$. Now for each pair $(\omega_j,\nu_\ell)$ in the Cartesian product of the above cited sets, A set of \texttt{n\_samples} \textbf{normalized} trajectories and their corresponding $d$-derivatives can be created:  
\begin{align}
X_{(j,\ell,d)} &:= \Biggl\{\dfrac{\texttt{B$_d$[:$n_w$,:$\omega_j$]}a^{[\kappa]}}{\|\texttt{B[:$n_w$,:$\omega_j$]}a^{[\kappa]}\|_\infty}+\delta_{0d}\cdot \nu_\ell\cdot \mathbf u^{[\kappa]}\Biggr\}_{\kappa\in \{1,\dots, \texttt{n\_samples}\}} \nonumber \\ 
&\in \Bigl[\mathbb R^{n_w}\Bigr]^{\texttt{n\_samples}}\label{defdeXjld}
\end{align} 
where for each $\kappa$, $a^{[\kappa]}$ is a random vector of dimension $n_B$ while $\mathbf u^{[\kappa]}$ is a random sequence of white noise of standard deviation=1. $\delta_{0d}$ is the Kronecker product that is equal to $0$ for all $d\neq 0$ while $\delta_{00}=1$. 
\e This means that only the $0$-derivative version is made noisy by adding the noise with the level $\nu_\ell$ to the normalized randomly generated time-series. 
\e This is done because the $0$-derivative version is the argument of the maps to be learned and that comes always corrupted by the measurement noise. By opposition, the set of $d$-derivatives are not corrupted with noise since they will serve as the ground truth labels for the model's fitting process. 
\e 
The previous discussion can be summarized as follows:
\begin{center}
\begin{tikzpicture}
\node[inner sep=5mm, fill=Black!5]{
\begin{minipage}{0.43\textwidth}
The lines of matrix $X_{(j,\ell,0)}\in \mathbb R^{\texttt{n\_samples}\times n_w }$ defined by \eqref{defdeXjld} represent a set of \texttt{n\_samples} time-series, each of amplitude lower than $1$ should the noise be $0$ and $X_{(j,\ell,d)}$ are their corresponding \textit{exact} $d$-derivatives time-series. For a given bandwidth index $j$ and a noise level index $\ell$, the base model for the reconstruction of the $d$-derivative for $d\ge 1$ can be learned using\footnote{  In Machine Learning terminology, the feature matrix is a matrix where each line is an instance of the input $x$ to the function $F$ we are looking for such that $F(x)$ equal the associated target, also called label, say $\ell$. Machine Learning algorithms try to approximately solve the set of equations:  $F(x^{(i)})\approx \ell_i$ for $i=1,\dots, n$ where $x^{(i)}$ is the $i$-th line of the features matrix having $n$ lines.}:\vskip 2mm
\begin{itemize}
\item the noisy $X_{(j,\ell,0)}$ \textbf{as features} matrix 
\item the ground-truth noise-free $X_{(j,\ell,d)}$ \textbf{as labels}
\end{itemize} 
\vskip 2mm
 As for the reconstruction of the $0$-derivative (filtering mode), the feature matrix and the label matrix are given by $X_{(j,\ell,0)}$ and $X_{(j,0,0)}$ respectively.
\end{minipage} 
};
\end{tikzpicture}
\end{center} 
So assume a features matrix $X\in \mathbb R^{\texttt{n\_samples}\times n_w }$ and a label matrix $L\in \mathbb R^{\texttt{n\_samples}\times n_w}$ (there is such a pair for each value of the triplet $(j,\ell,d)$), the model we are looking for is a matrix $A\in \mathbb R^{n_w\times n_w}$ that solves a regularized least-squares problem of the form:
\begin{equation}
\min_A \left\|XA-L\right\|^2+\alpha \|A\|^2 \label{optprob}
\end{equation} 
for an appropriate choice of the regularization parameter $\alpha>0$. The role of the regularization parameter $\alpha$ is to precisely achieve a trade-of between the need for capturing the relationship we are looking for and capturing the specific realization of the noise as invoked earlier. 
\e It is precisely in order to find this \textit{appropriate choice} for each triplet $(j,\ell,d)$ that the second idea is invoked:
\\ 
\begin{center}
\begin{tikzpicture}
\node[rounded corners, fill=Black!5, inner sep=4mm](O){
\begin{minipage}{0.43\textwidth}
\color{Black}
The fine tuning of the regularization parameter $\alpha$ involved in the optimization problem \eqref{optprob} can be automatically obtained  using the cross-validation technique widely used in the Machine Learning algorithms.
\end{minipage} 
};
\node[above, anchor=south west] at(O.north west){\sc Idea 2: $\alpha$ is  tuned via cross-validation};
\end{tikzpicture}
\end{center} 
Discussion. {\textit{The cross-validation technique consists in splitting the available learning data into \texttt{cv} (=2 hereafter) subsets of data. The model is learned (for a given regularization parameter value $\alpha$) using \texttt{cv-1} subsets leaving one subset for the evaluation of the extrapolation error on unseen data. This is repeated \texttt{cv} times and the statistics of the extrapolation error is computed for this specific $\alpha$. The finally chosen $\alpha$ is the one that minimizes the so-computed extrapolation error on unseen data.}
\e 
\textbf{Implementation.} For linear models as the ones we are seeking here, the Machine Learning libraries offer an already implemented fitting algorithms with a cross-validation-based auto-tuning of regularization parameters. In particular this is the case for the \texttt{scikit-learn} library \cite{scikit-learn} through the \texttt{RidgeCV} (used hereafter) and the \texttt{LassoLarsCV} models to cite but two examples. A typical \texttt{python} call involving the features and label matrices $X$ and $L$ takes the following form:\vskip 2mm
{
\hrule
\vskip 2mm 
\begin{verbatim}
from sklearn.linear_model import RidgeCV 
alphas = np.logspace(-4,3,20)
reg = RidgeCV(cv=2, alphas=alphas, 
            fit_intercept=False).fit(X, L)
\end{verbatim}
}
\hrule 

\e 
where \texttt{alphas} is a list of candidate values for the regularization parameter $\alpha$. The function \texttt{RidgeCV} performs the  above described cross-validation based optimization and return the appropriate matrix $A= \texttt{reg.coef\_} $ as an attribute of the fitted model \texttt{reg}. 
\e Notice that for the sake of simplification, we  omitted the reference to the triplet of indexes $(j,\ell,d)$ that underlines each of the associated solution $A$ which results, each, from its own pair of features and label matrices as explained above. \e As a matter of fact, upon exploring all possible values of the triplet $(j,\ell,d)$, a dictionary of models (matrices) is obtained such that:
\begin{equation}
\Bigl\{A_{(j,\ell,d)}\in \mathbb R^{n_w\times n_w}\Bigr\}_{(j,\ell,d)\in \{1,\dots,n_r\}\times\{1,\dots, q\}\times\{0,1,\dots,d_{max}\}}
\end{equation} 
Recall that $j$ denotes the index of the maximum pulsation $\omega_j\in \bar\Omega_\texttt{design}$ [see \eqref{defdeOmegadesign}] contained in the time-series while $\ell$ denotes the noise level index associated to the standard deviation $\nu_\ell$ defined in \eqref{noiselevels}. We shall later explain how these appropriate indices are computed for a given time-series but let us before explain the main averaging idea that is in the heart of the precision improvement provided by the framework.
\e In order to explain the idea let us assume that the triplet $(j,\ell,d)$ is available and hence so is its associated model summarized by the matrix $A_{(j,\ell,d)}\in \mathbb R^{n_w\times n_w}$ such that for any time series $\mathbf y\in \mathbb R^{n_w}$ of bandwidth $\omega_j$ and noise level $\nu_\ell$, the estimation of its $d$-derivative time-series is given by:
\begin{equation}
\hat{\mathbf y}^{(d)}:= \bigl[A_{(j,\ell,d)}\bigr]^T\cdot \mathbf y
\end{equation} 
let us consider a \textbf{longer time series} $\mathbf s\in \mathbb R^{n}$ with $n\gg n_w$. Notice that each slice of $\mathbf s$ of length $n_w$, namely $\mathbf s[i:i+n_w-1]$ for some $i\le n-n_w+1$ can be viewed as a time-series $\mathbf y_{[i]}(\mathbf s)\in \mathbb R^{n_w}$ for which one can use the models $A_{(j,\ell,d)}$ to reconstruct the $d$-derivative via: 
\begin{align}
E^{(j,\ell,d)}(\mathbf y_{[i]}(\mathbf s))=:\begin{bmatrix}
E^{(j,\ell,d)}_0(\mathbf y_{[i]}(\mathbf s))  \cr \vdots \cr 
E^{(j,\ell,d)}_{n_w-1}(\mathbf y_{[i]}(\mathbf s))
\end{bmatrix}:=\bigl[A_{(j,\ell,d)}\bigr]^T\cdot \underbrace{\mathbf s[i:i+n_w-1]}_{\mathbf y_{[i]}(\mathbf s)}\label{letruccomplique}
\end{align} 
where $j$ and $\ell$ corresponds to the bandwidth and the noise level of the time-series $\mathbf s$. Before we state the next averaging idea, let us summarize the last \textit{rather complicated notation} as follows:
\begin{center}
\begin{tikzpicture}
\node[fill=Black!5, inner sep=4mm]{
\begin{minipage}{0.43\textwidth}
Given a time-series $\mathbf s\in \mathbb R^{n}$ where $n\ge n_w$ that corresponds to a bandwidth $\omega_j$ and noise-level close to $\nu_\ell$, the term $E^{(j,\ell,d)}_k(\mathbf y_{[i]}(s))$ defined by \eqref{letruccomplique} provides an estimation of the $d$-derivative of $\mathbf s$ at instant $i+k$. \e Consequently, given any $m\in \{1,\dots,n\}$, all pairs in the set of pairs defined by $$\mathcal I_m^{(n,n_w)}:=\Bigl\{(i,k)\in \{1,\dots,n\}\times \{0,\dots, n_w-1\}\ \vert\ i+k=m\Bigr\}$$
provide as many \textbf{elligible} estimations of the $d$-derivative of $\mathbf s$ \textsc{at the same instant $m$} as there are elements in $I_m^{(n,n_w)}$, namely \texttt{card}$(I_m^{(n,n_w)})$.
\end{minipage} 
};
\end{tikzpicture}
\end{center} 
\noindent But it is easy to figure out that when $m$ spans the set of indices of the time series $\mathbf s$, namely $\{1,\dots,n\}$, the number of estimations that can be gathered is given by:
\begin{equation}
\texttt{card}(I_m^{(n,n_w)}):= \left\{\begin{array}{ll}
m & \text{if $m < n_w$} \cr 
n_w & \text{if $n_w\le m\le n-n_w$} \cr 
n-m & \text{if $m >  n-n_w$} 
\end{array}
\right.
\end{equation} 
This means that when $n\gg n_w$, $n_w$ estimations are obtained for the majority of time instants. This leads to the statement of the next idea:
\begin{center}
\begin{tikzpicture}
\node[rounded corners, fill=Black!5, inner sep=4mm](O){
\begin{minipage}{0.45\textwidth}
\color{Black}
By using high values of $n_w$ and by considering long time-series $\mathbf s$, it is possible to get two benefits, namely:\vskip 3mm
\begin{enumerate}
    \item Reducing the impact of noise by averaging multiple estimations of the $d$-th derivative for each single instant; \vskip 2mm
\item Getting confidence intervals of the reconstruction by measuring the dispersion of the different estimated values at each single instant.
\end{enumerate}
\end{minipage} 
};
\node[above, anchor=south west] at(O.north west){\sc Idea 3: Reducing noise impact via moving window averaging};
\end{tikzpicture}
\end{center} 
Discussion. {\textit{There are obviously some limitations on the use of larger values for $n$ and $n_w$. Indeed, regarding $n_w$, one should keep in mind that for each triplet of values $(j,\ell,d)$ the base model involves a model of size $n_w^2$ (the number of elements in the matrix $A_{(j,\ell,d)}$) and all these models should be stored for possible use. Notice however that the memory can be drastically reduced using different possible compression techniques including the Singular Value Decomposition (SVD) technique that is used in the model explored later in this paper. in other words, all the results shown later are based on compressed models in order to check that the performances are those of the compressed models that should be ultimately used to get a light differentiation portable package. As for the limitation of the length $n$ of the analyzed time-series, it stems from the risk of having \textit{non differentiable incidents} that might impact the quality of the estimation of the bandwidth of the signal. Moreover, even in the absence of such incidents, the associated risk is to consider high bandwidth pulsation $\omega_j$ that lasts only over a small portion of the window reducing the quality of the estimation on low frequency parts of the window.}  
}
\e 
In all the presented results, the implementation uses $n_w=50$.  The higher values of $n_w=100, 200, 400$ have been tested showing slightly higher precision scores but the increment does not necessarily justify the impact on the memory footprint of the resulting dictionary of  maps. {  Using $n_w=50$, the resulting size of the cumulative memory of all the compressed models for all possible triplets put together is around 15Mb}. 
\e 
\textbf{Implementation}. Based on the above discussion, given a time-series $\mathbf s\in \mathbb R^{n}$, the estimation of the $d$-derivative at instant $m$ is given by:
\begin{equation}
\hat{\mathbf s}^{[d]}_m:= \frac{1}{\texttt{card}(I_m^{(n,n_w)})}\sum_{(i,k)\in I_m^{(n,n_w)}}E_k^{(j,\ell,d)}(\mathbf s[i:i+n_w-1])
\end{equation} 
or in matrix form, denoting by $M[k,:]$ the $k$-th line of a matrix $M$:
\begin{equation}
\hat{\mathbf s}^{[d]}_m:= \frac{1}{\texttt{card}(I_m^{(n,n_w)})}\sum_{(i,k)\in I_m^{(n,n_w)}}A_{(j,\ell,d)}^T[k,:]\cdot \mathbf s[i:i+n_w-1]\label{shatformulae}
\end{equation} 
Similarly, the standard deviation of the estimation can also be obtained according to:
\begin{equation}
\hat{\mathbf \sigma}^{[d]}_m:= \Biggl[\frac{1}{\texttt{card}(I_m^{(n,n_w)})}\sum_{(i,k)\in I_m^{(n,n_w)}}\left\vert E_k^{(j,\ell,d)}(\mathbf s[i:i+n_w-1])-\hat{\mathbf s}^{[d]}_m\right\vert^2\Biggr]^{\frac{1}{2}}\label{compsigma}
\end{equation} 
which can also be put in matrix form. 
\e This \textit{almost} achieves the target as stated in Section \ref{sec:pbstat} through \eqref{targetedmap}. \textit{Almost} because we considered that the bandwidth index  $j$ and the noise level index $\ell$ are available so that the appropriate model's matrix $A_{(j,\ell,d)}$ can be selected and used. The ways these indices are determined for a time-series $\mathbf s$ are successively described in the remainder of this section. 
\e 
\subsection{bandwidth index selection}\label{sec:bandwidthIndexSelection}
\noindent Recall that the appropriate index $j$ we are looking for is the minimum index $j\in \{1,\dots,n_r\}$ such that the time-series $\mathbf s$ under scrutiny contains no pulsations that are greater than $\omega_j\in \bar\Omega_\texttt{design}$. The selection is based on the behavior of the error between the noisy signal $\mathbf s$ and its projection on the sub-space generated by the columns of the $\omega_j$-truncated matrix defined by \eqref{defdeBpointpointomega}-\ref{defdePin} when $j$ increases from $1$ to $n_r$. The error is computed using the projection matrix $\Pi_n(\omega_j)$ defined by \eqref{defdePin} according to:
\begin{equation}
e_j:= \Bigl\|(\Pi_n(\omega_j)-\mathbb I_n)\cdot \mathbf s\Bigr\|
\end{equation} 
The appropriate value $j^\star$ is obtained when the decrease of the error $e_j$ becomes negligible meaning that adding higher pulsations (more columns) does not improve the approximation. More precisely the following selection rule is used:
\begin{equation}
j^\star:=\inf \Bigl\{j\in \{1,\dots,n_r\}\quad\vert\quad (e_j-e_{n_r})\le \texttt{th}\times(e_1-e_{n_r})\Bigr\}\label{defdejstar}
\end{equation} 
where \texttt{th}$>0$ is a small threshold (\texttt{th}=0.1 is used in the implementation).
\subsection{Estimating the noise level}
\noindent Despite the fact that the level of the noise affecting the measured signal can be roughly guessed from a simple inspection of a given signal, the need for a systematic blind solution stems from the fact that a visual inspection step might not be allowed. Moreover, the level of noise can vary along large datasets that can span months if not years of data collection on the other hand. During such long periods, the sensors might be changed and/or deteriorated over time. The conclusion is that some systematic algorithmic correction should be considered. The way this can be done is explained in this section. 
\e In order to understand the proposed solution, an important fact should be first stated: The precise knowledge of the noise level lead to a second order improvement of the quality of the reconstructed derivatives. This means that even if a model with \textit{erroneous} noise level is used, the quality of the estimation of the $0$-derivative  of the signal (filtered version of the original signal) is still sufficiently good to get a much better estimation of the truly involved noise level. Once this estimation is available, a second call of the appropriate model can be done to get even better result. 
\e 
This leads to the solution involving the following steps:
\begin{enumerate}
\item First determine the bandwidth index $j^\star$ as explained in the previous section [see \eqref{defdejstar}].\\
\item  If available, use the user-provided value of the noise level. If no such a knowledge is available, use $\ell$ such that $\nu_\ell=0.05$.\\
\item Compute the filtered version of the time-series $\mathbf s$ using the $(j^\star,\ell,0)$ model, namely:
\begin{equation}
\mathbf s_f := \Bigl[A_{(j^\star,\ell,0)}\Bigr]^T\cdot \mathbf s 
\end{equation} 
\item Estimate the standard deviation of the noise according to:
\begin{equation}
\sigma^\star := \texttt{std}\Bigl[\mathbf s-\mathbf s_f\Bigr] 
\end{equation} 
\item Find the noise level index $\ell^\star$ such that:
\begin{equation}
\ell^\star := \texttt{arg}\min_{\ell=1}^{n_r} \vert \nu_\ell-\sigma^\star\vert 
\end{equation} 
\item Use the model indexed by $(j^\star, \ell^\star, d)$ to compute the desired $d$-derivative of the time-series $\mathbf s$:
\begin{equation}
\hat{\mathbf s}^{[d]} := \Bigl[A_{(j^\star,\ell^\star,d)}\Bigr]^T\cdot \mathbf s 
\end{equation} 
\end{enumerate}
This ends the presentation of the proposed framework. It is now time to proceed to some numerical assessments and comparisons. This is the aim of the next section. 
\section{Assessment through numerical investigations}\label{secs:assessment}
\noindent Before we show comparison with alternative solutions, let us first examine typical derivative reconstruction results that can be obtained using the proposed framework. This is shown in Figure \ref{fig_valid_der_4}. In this figure, the yellow regions represent the $3\sigma$-confidence intervals computed according to \eqref{compsigma}. The quality of the averaging-based reconstruction formulae \eqref{shatformulae} compared to the width of the confidence zone assesses the crucial role played by the moving window averaging in the quality of the derivative reconstruction. 
\e The performance of the proposed algorithm is compared to the performance of the competing algorithms as described in the following section. For each algorithm, the tuning parameters are described and their ranges used in the forthcoming results are given. Regarding the settings of the parameters of the alternative methods, the following approach is used which is extremely favorable to all the algorithms except the proposed one:
\begin{center}
 \begin{tikzpicture}
 \node[rounded corners, fill=Black!5, inner ysep=6mm, inner xsep=3mm, draw=Black](T){
\begin{minipage}{0.43\textwidth}
For each experiment and for each derivation order, the best tuning of the alternative methods (among the below defined set of tuning parameters values for each method) is chosen based on the ground truth of targeted derivatives. This leads to a favorable comparison for the alternative solutions {  since the ground truth is not supposed to be known and hence the associated optimal setting is impossible to guess by this means}.
\end{minipage}
};    
\node[fill=Black, rounded corners] at(T.north){\color{white}\footnotesize Tuning rule for alternative solutions};
 \end{tikzpicture}
\end{center}
This leads, for the alternative algorithms, to a different ground-truth based setting, for each experiment and each derivative order to reconstruct inside the same experiment. 
% Moreover, the domain of variation and the structure of the parameterization have been chosen after some hand-made optimization round in order to produce the best possible results for each of the  alternative algorithms that are described in the next section. 
\subsection{Alternative algorithms}
\noindent The following derivatives estimation methodologies are examined in this section:
\e 1) \textbf{Kalman Filter}: This is a well known and widely used option that is based on building a state observer for the dynamic system given by: 
\begin{equation}
\dot x_i=x_{i+1} \qquad i\in \{1,\dots,d_{max}\} \quad \dot x_{d_{max}+1}=0\label{eqkalman}
\end{equation}
when the estimation of the derivatives up to order $d_{max}$ is required. The design of this filter needs the weighting matrices $Q$ and $R$ on the state and the measurement to be provided. To this end, the following parameterization is used: 
\begin{equation}
R=1\quad;\quad Q:= \nu_s\times \texttt{diag}\Bigl(\rho,\dots,\rho^{d+1}\Bigr) \label{}
\end{equation}
where the two dimensional parameter $p_\text{kalman}:=(\nu_s,\rho)$ takes possible tuning values in the set:
\begin{equation}
\small 
\mathbb P_\text{kalman}:= \texttt{logspace(-21,21,25)} \times \texttt{logspace(0,8,10)} \label{defdePkalman}
\end{equation}
which encompasses 250 different settings.
\e 
2) \textbf{Spectral derivation}. This algorithm is based on the property ($\mathcal F$ denoting the fourier transform):
\begin{equation}
\mathcal F[y^{(d)}](s) = s^d \times \mathcal F[y](s) \label{fourierderivative}
\end{equation}
which enables to estimate the $d$-derivative by applying the inverse Fourier transform to the multiplication, in the frequency domain, by $s^d$ of the Fourier transform of the noisy original time series. The tuning parameter for this approach is the \textit{smoothing} filter applied to the original signal before to process it as described above. In the implementation used hereafter, a Gaussian filter is used of the form $G(j\omega):=\exp(-\mu_f\cdot \omega^2)$ in which the smoothing parameter $p_\text{spectral}:=\mu_f$ takes values in the tuning set given by:
\begin{equation}
\mathbb P_\text{spectral} := \texttt{logspace(-6,0,50)} \label{defdePspectral}
\end{equation}
leading to 50 different settings.\e 
3) The \texttt{scipy} \textbf{Savitzky-Golay} filter. This filter \cite{krishnan2012selection} is based on an iterative window polynomial smoothing of the time series which result in two parameters: the window size $n_w$ and the order of the polynomial $r$ with the condition $r<n$. The following admissible sets are used for these two parameters: 
\begin{align*}
\mathcal N_w&:=\Bigl\{1, 5, 11, 21, 41, 51, 101, 201, 401, 501\Bigr\} \\
\mathcal R&:=\{2,3,4,5\}
\end{align*}
leading to the following set of $40$ possible settings of the parameter $p_\text{savgol}:=(n_w, r)$:
\begin{equation}
\mathbb P_\text{savgol}:= \mathcal N_w \times \mathcal R \label{}
\end{equation} 
3) {  The Implicit \textbf{AO-STD filter}.  This filter promoted in \cite{mojallizadeh2021time} implements the implicit version of a sliding mode filter which can provide any order derivatives. It takes the following form in in which $z_{i,k}$ stands for the estimated $i$-th derivative at instant $k$:
\begin{align}
z_{n,k+1}&=-h\lambda_nL \texttt{sign}(\sigma_{0,k+1})+z_{n,k} \label{brogli_1}\\
z_{i,k+1}&=-h\lambda_i L^{\frac{i+1}{n+1}}\vert \sigma_{0,k+1}\vert^{\frac{n-i}{n+1}}\texttt{sign}(\sigma_{0,k+1})+h z_{i+1,k+1}+z_{i,k}\label{brogli_2}
\end{align}
which is an implicit equation in $\sigma_{0,k+1}:= z_{0,k+1}-y_{k+1}$ representing the next estimation error on the original noisy signal $y$ at the next instant $k+1$. The implementation uses exact implicit inversion via fixed-point iteration and the parameter $\lambda_i$ suggested in Table 1 of \cite{mojallizadeh2021time}. This leaves us with the hyper parameter $L$ that is taken in the set defined by \texttt{logspace(-6,6,100)} which contains hundred possible settings the best of which is taken for every single profiles in the benchmark which is tremendously favorable to the competing solutions. 
}
\subsection{The benchmark}
\noindent A set of validation time-series that are randomly generated using a set of pairs $(\omega,\nu)$ of bandwidths and noise levels that belong to the Cartesian product of the sets:
\begin{subequations}
\begin{align}
\mathbb W&:=\Bigl\{0.01, 0.05, 0.1, 0.2, \dots,0.8, 0.9, 0.95\Bigr\} \label{defdeBBW}\\
\mathbb L&:=\Bigl\{0.01, 0.02, 0.03, 0.04, 0.05, 0.06, 0.07, 0.1\Bigr\}\label{defdeBBL}
\end{align}
\end{subequations}
\noindent leading to a 96 time series of length 2000 each which corresponds to a total number of instants equal to 192000. 
\e As for the performance evaluation, the following definition of the $d$-th derivative's reconstruction error is used:
\begin{equation}
e:=\dfrac{\texttt{percentile}\Bigl(\vert \hat{\mathbf s}^{(d)}-\mathbf s^{(d)\vert}\vert, 95\Bigr)}{\texttt{percentile}(\vert \mathbf s^{(d)}\vert, 50)} \label{deferror}, 
\end{equation}
in which $\hat{\mathbf s}^{(d)}$ and $\mathbf{s}^{(d)}$ stand for the reconstructed and the true $d$ derivatives.
\e Figure \ref{fig_percentiles} shows the comparison of the error as defined above over the whole validation dataset for the different derivative reconstruction algorithms described above. This figure suggests that the spectral and Savitzky-Golay options are very close to each other (as far as the adopted optimistic tuning rule is concerned) but they are both quite largely outperformed by the proposed algorithm and this is increasingly obvious as the degree of the derivation is increased. \e  Moreover, it is shown in Figure \ref{fig_dependence_on_noise} that the quality of the estimation under these algorithms continuously deteriorates as the derivation order increases while this quality seems steady when using the proposed algorithm as the derivation order increases. In order to assess the relevance of the confidence intervals, Table \ref{tab1} shows the ratio (over the 192,000 instants) of those where the reconstruction errors are lower than different thresholds: $\sigma/2$, $\sigma$, $2\sigma$ and $3\sigma$.
\e {  Regarding the computation time, notice that the current implementation enables to compute the derivatives of a time-series of length 100 in less than 100 msec. This makes the approach eligible for real-time on-line implementation  for a class of systems with comparable characteristic times. }
\begin{table}
\begin{center}
\footnotesize
\begin{tabular}{lrrrr}
\hline
 & $\mathbf{d=1}$ & $\mathbf{d=2}$ & $\mathbf{d=3}$ & $\mathbf{d=4}$ \\
\hline
$\sigma$/2 & 0.35 & 0.48 & 0.40 & 0.45 \\
$\sigma$ & 0.61 & 0.73 & 0.65 & 0.70 \\
2$\sigma$ & 0.87 & 0.92 & 0.87 & 0.91 \\
3$\sigma$ & 0.95 & 0.97 & 0.95 & 0.97 \\
\hline
\end{tabular}
\end{center}
\caption{\textbf{Relevance of the confidence intervals}: Ratios of the instants where the reconstruction error is within the thresholds $\sigma/2$, $\sigma$, 2$\sigma$ or $3\sigma$.}\label{tab1}
\end{table}
% \begin{figure}
% \begin{center}
% \includegraphics[width=0.38\textwidth]{images/valid_der_6.png}  
% \end{center}
% \caption{Typical results of derivative reconstruction up to order 4 under noisy measurements with standard deviation of 7\%. Raw signal of bandwidth 18 Hz. Yellow bands show the 3$\sigma$-confidence intervals.}\label{fig_valid_der_6}
% \end{figure}
\begin{figure}
\begin{center}
\includegraphics[width=0.38\textwidth]{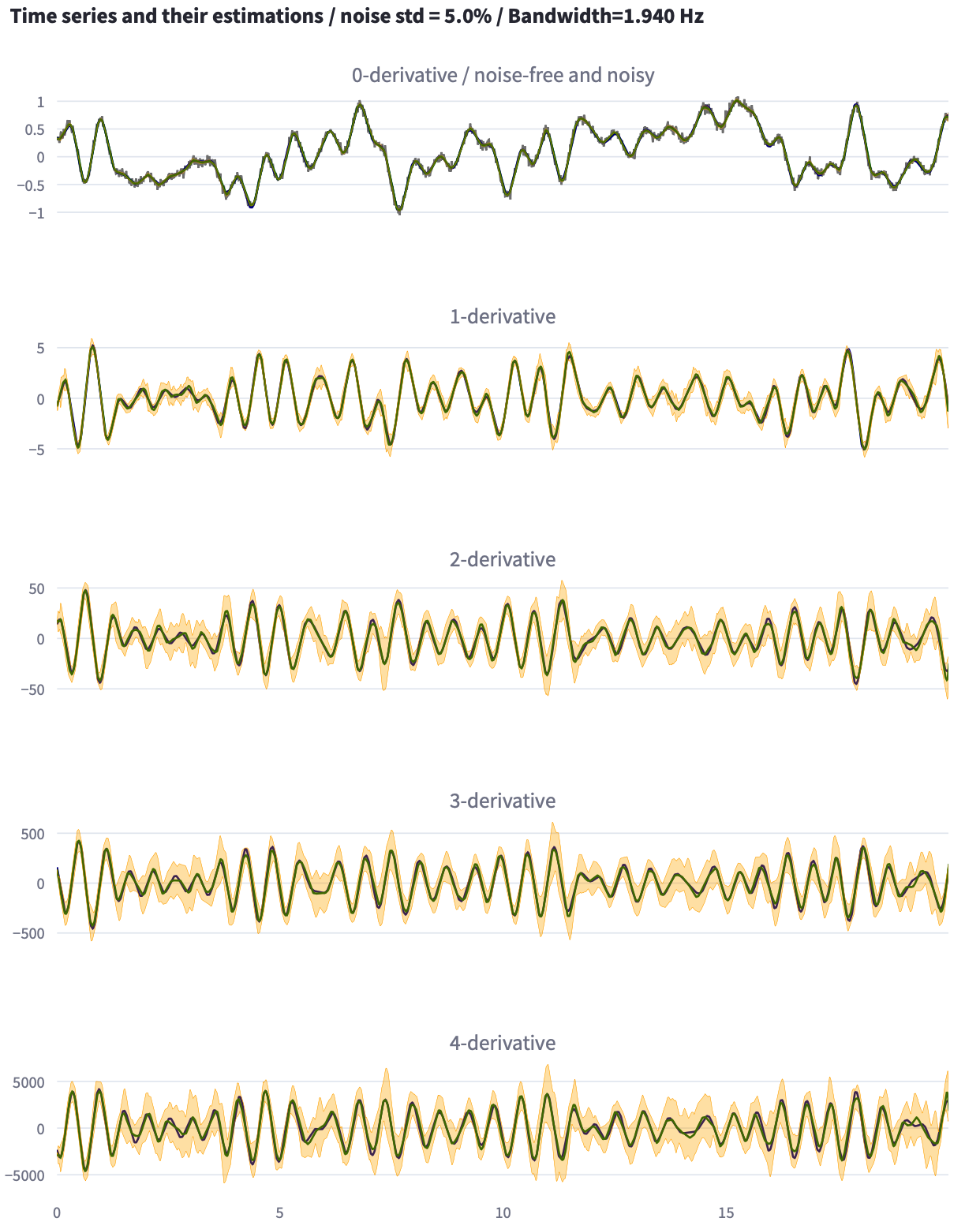}  
\end{center}
\caption{Typical results of derivative reconstruction up to order 4 under noisy measurements with standard deviation of 5\%. Raw signal of bandwidth 1.94 Hz. Yellow bands show the 3$\sigma$-confidence intervals. {  Notice that both exact and estimated values are plotted.}}\label{fig_valid_der_4}
\end{figure}
\begin{figure}
    \centering
    \includegraphics[width=0.9\linewidth]{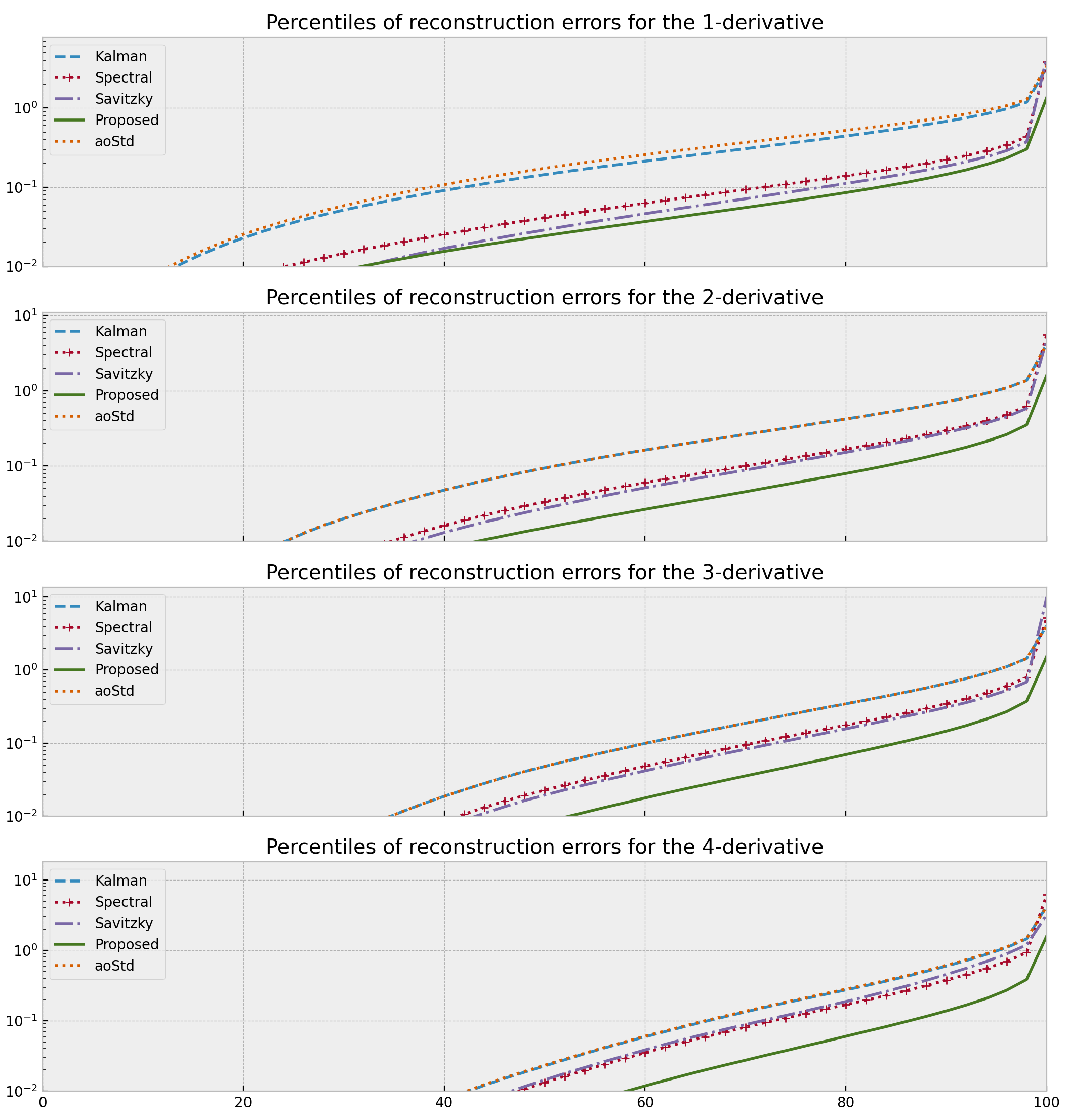}
    \caption{  Percentiles of reconstruction errors on the first four derivatives for the different algorithms (logarithmic scale). These statistics are computed over all the scenarios including the ones where a standard deviation of 10\% is used for the noise. The total number of samples is equal to 192000. The $y$-scale is lower bounded by $10^{-2}$ for a better readability of the comparison. The best parameters is used based on the ground truth for each reconstruction of the alternative solutions.}
    \label{fig_percentiles}
\end{figure}
\begin{figure}
    \centering
    \includegraphics[width=0.9\linewidth]{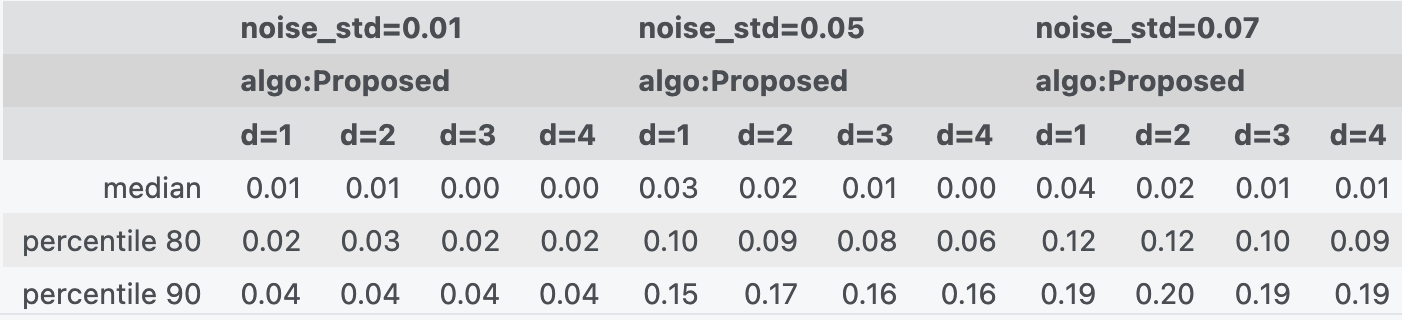}
    \includegraphics[width=0.9\linewidth]{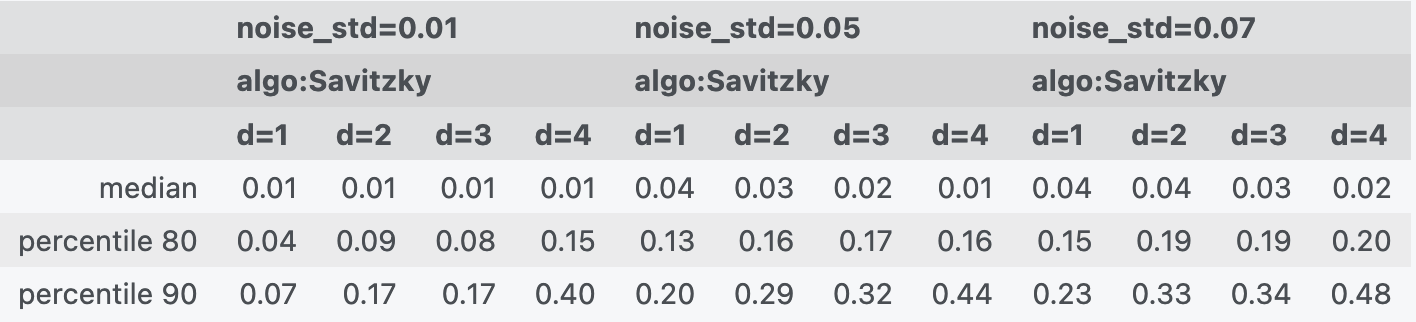}
    \includegraphics[width=0.9\linewidth]{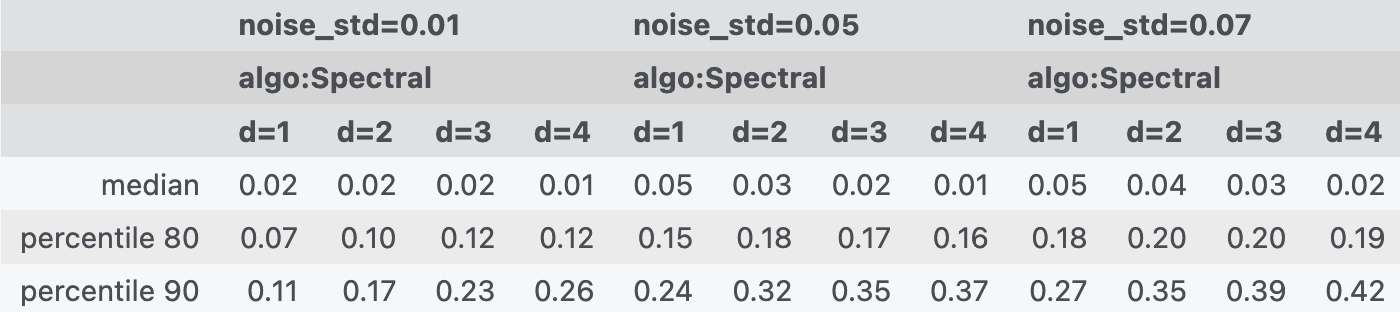}
    \includegraphics[width=0.9\linewidth]{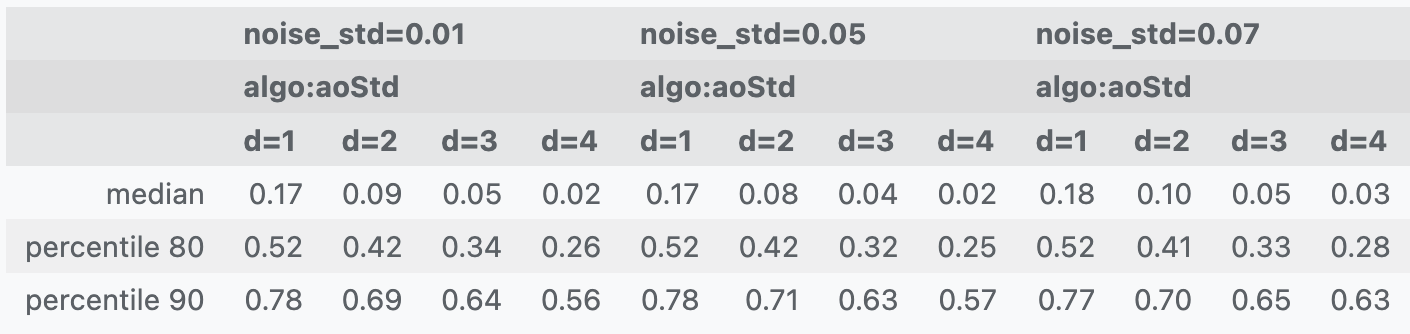}
    \caption{  Evolution of the statistics of the error as the measurement noise increases for the different algorithms. The best parameters is used based on the ground truth for each reconstruction of the alternative solutions.}
    \label{fig_dependence_on_noise}
\end{figure}
\section{Conclusion}
\noindent In this paper a new tuning-free algorithm for the reconstruction of high derivatives of noisy time-series is proposed. The algorithm provides, in addition to derivatives reconstruction, a consistent confidence interval that can be used in the selection of the windows over which the results can be kept for later use in the identification and/or characterization of the normality among many possible tasks.

\bibliographystyle{elsarticle-harv}  % Elsevier Harvard style for references
\bibliography{bib_derivation}            % Path to your .bib file

\end{document}